\documentclass[twocolumn,showpacs,nofootinbib,tightenlines,nobibnotes,aps,amssymb,amsmath,prb]{revtex4}
\usepackage{graphicx}

\begin{document}
\title{Quantum critical dynamics of the two-dimensional Bose gas}
\author{Subir Sachdev and Emily R. Dunkel}
\affiliation{Department of Physics, Harvard University, Cambridge,
MA 02138}

\date{October 11, 2005}

\begin{abstract}
The dilute, two-dimensional Bose gas exhibits a novel regime of
relaxational dynamics in the regime $k_B T \gtrsim |\mu|$ where
$T$ is the absolute temperature and $\mu$ is the chemical
potential. This may also be interpreted as the quantum criticality
of the zero density quantum critical point at $\mu=0$. We present
a theory for this dynamics, to leading order in $1/\ln (\Lambda/
(k_B T))$, where $\Lambda$ is a high energy cutoff. Although
pairwise interactions between the bosons are weak at low energy
scales, the collective dynamics are strongly coupled even when
$\ln (\Lambda/T)$ is large. We argue that the strong-coupling
effects can be isolated in an effective classical model, which is
then solved numerically. Applications to experiments on the
gap-closing transition of spin gap antiferromagnets in an applied
field are presented.
\end{abstract}


\maketitle

\section{Introduction}

Despite the widespread recent theoretical and experimental
interest in quantum phase transitions, a direct quantitative
confrontation between theory and experiment has been difficult to
achieve for systems in two and higher spatial dimensions. A major
obstacle is that it is often difficult to tune system parameters
over the range necessary to move across a quantum critical point.
Furthermore, for many examples where such tuning is possible, the
theory for the quantum critical point is intractable.
Consequently, the analysis of the data is often limited to the
testing of general scaling ansatzes, without specific quantitative
theoretical predictions.

A class of quantum phase transitions have recently been
exceptionally well characterized in a variety of experiments.
These experiments study the influence of a strong applied magnetic
field on insulating spin-gap
compounds\cite{oosawa,sebastian,zapf,stone,stone2,hong}. The low
lying spin excitations behave like spin $S_z=1$ canonical Bose
particles, and the energy required to create these bosons vanishes
at a critical field $H=H_c$, which signals the position of a
quantum phase transition\cite{sss} with dynamic critical exponent
$z=2$. In spatial dimensions $d=3$, the quantum critical
fluctuations are well described by the familiar Bose-Einstein
theory of non-interacting bosons, and no sophisticated theory of
quantum criticality is therefore necessary to interpret the
experiments. The upper-critical dimension of the quantum critical
point is $d=2$, and the boson-boson interaction vanishes
logarithmically at low momenta. So naively, one expects that the
$d=2$ case is also weakly coupled, and no non-trivial quantum
critical behavior obtains.

The primary purpose of this paper is to show that the above
expectation for the quantum-criticality of $d=2$ Bose gas is
incorrect. While the pairwise interactions between the bosons are
indeed weak, the collective properties of the finite-density,
thermally excited Bose gas pose a strong-coupling problem. We will
demonstrate here that an effective classical model, which can be
readily numerically simulated, provides a controlled description
of this problem. This will allow us to present predictions for the
evolution of the dynamic spectrum of the $d=2$ Bose gas across the
quantum critical point at non-zero temperatures.

Our results can be applied to two-dimensional spin-gap
antiferromagnets, in the vicinity of the gap-closing transition
induced by an applied magnetic. Recent
experiments\cite{stone,stone2,hong} on piperazinium
hexachlorodicuprate (PHCC), (C$_4$H$_{12}$N$_2$)Cu$_2$Cl$_6$, will
be compared with our results in Section~\ref{sec:conc}. These
experiments are able to easily access the finite temperature
quantum-critical region, which we will describe by our theory of
the dilute Bose gas in $d=2$; this Bose gas is in a regime of
parameters which is typically not examined in the context of more
conventional atomic Bose gas systems \cite{popov,fisher}.

Results similar to those presented here \cite{ashvin} also apply
to other quantum critical points with upper-critical spatial
dimension $d=2$ {\em i.e.\/} while the zero temperature properties
can be described in a weak-coupling theory, all non-zero
temperature observables require solution of a strong-coupling
problem. A prominent example of such a quantum-critical point is
the Hertz theory of the onset of spin-density-wave order in a
Fermi liquid\cite{hertz}. Results for the thermodynamic and
transport properties can be obtained by the methods presented here
\cite{ashvin}. We maintain that such results are necessary for
understanding experiments, and that previous theoretical
results\cite{hertz,moriya,millis} are inadequate for a
quantitative analysis.

\begin{figure}
\centering
\includegraphics[width=3.2in]{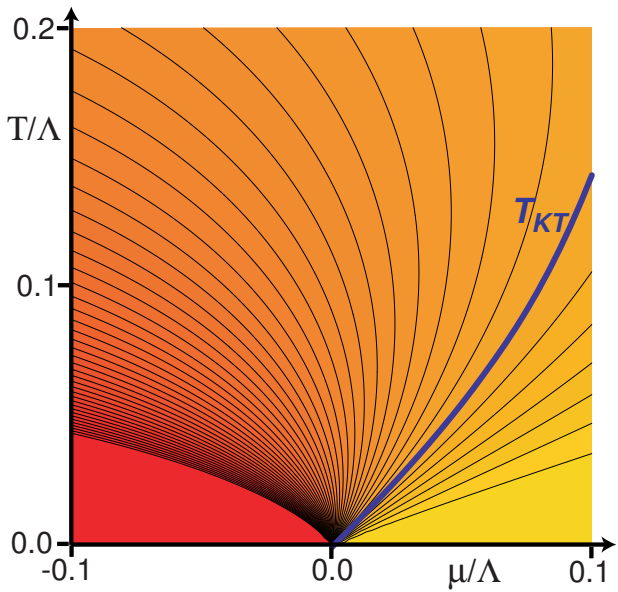}
\caption{Color density plot of the value of the dimensionless
ratio $U/R$ as a function of the chemical potential $\mu$ and the
temperature $T$ of a dilute Bose gas in two spatial dimensions.
The energy scale $U$, defined in Eq.~(\ref{defU}), is a measure of
the pairwise interaction between the atoms, while $R$, defined in
Eq.~(\ref{defR}), is an energy scale controlling collective
excitations. The physical properties of systems with the same
value of $U/R$ are the same, apart from the change in the value of
the energy scale $R$. The Kosterlitz-Thouless transition occurs at
the contour labelled $T_{KT}$ at \cite{ss1,nikolay1} $U/R \approx
34$, and the superfluid phase is present for larger $U/R$. The
contours shown are for equal spacings of values of $R/U$, with the
yellow region on the lower right including values $R/U \rightarrow
0$. Our primary results here are in the quantum critical region,
with intermediate values of $U/R$ at $\mu=0$ and $T > 0$.}
\label{fig1}
\end{figure}

Our results will be presented in the context of the phase diagram
of the $d=2$ Bose gas shown in Fig.~\ref{fig1} as a function of
the boson chemical potential, $\mu$, and temperature, $T$.
At $T=0$, there is a quantum critical point at $\mu=0$. For $\mu <
0$, the ground state is simply the vacuum with no bosons, while
for $\mu > 0$, there is a finite density of bosons in the ground
state. In the spin gap antiferromagnets, $\mu = g \mu_B (H -
H_c)$, where $H$ is the applied field, $H_c$ is the critical
field, $g$ is the gyromagnetic ratio, and $\mu_B$ is the Bohr
magneton.

Fig.~\ref{fig1} presents results in terms of contours with
equivalent physical properties, up to an overall energy scale
($R$) whose $T$ and $\mu$ dependence will be explicitly presented
below, as will the equations determining the shape of the
contours. One of the contours is the position of the
Kosterlitz-Thouless (KT) transition of the $d=2$ Bose gas, which
is present only for $\mu > 0$. We are primarily interested here in
the quantum-critical region,\cite{sss} which is roughly the region
with $k_B T \gtrsim |\mu|$.  In particular, we will be able to
quantitatively examine the signature relaxation rate $\sim k_B T
/\hbar$ of the quantum-critical regimes: the ``Bose molasses''
dynamics. Our results and methods also describe the crossover to
the KT transition, as well as (in principle) the region with $T<
T_{\rm KT}$.

\subsection{Summary of results}
\label{sec:summary}

We summarize here the universal aspects of our results of the
quantum-critical Bose dynamics in two spatial dimensions. The
continuum quantum field theory of the critical point has
logarithmic corrections to scaling; consequently, the properties
of the continuum theory do have a logarithmic dependence upon a
non-universal ultraviolet cutoff. We will show that this cutoff
dependence can be isolated within a single parameter; all other
aspects of the theory remain universal, and can be accurately
computed.

We will be interested in the continuum Bose gas theory with the
partition function
\begin{eqnarray}
\mathcal{Z}_B &=& \int \mathcal{D} \psi ({\bf r}, \tau)
e^{-\mathcal{S}_B/\hbar} \nonumber \\
\mathcal{S}_B &=& \int_0^{\hbar /k_B T} d \tau \int d^2 r \Biggl[
\hbar \psi^{\ast} \frac{\partial \psi}{\partial \tau} +
\frac{\hbar^2}{2m} \left| \nabla_{\bf r} \psi \right|^2 - \mu
|\psi|^2 \nonumber \\
&~&~~~~~~~~~~~~~~~~~~~~~~~+ \frac{V_0}{2} |\psi |^4  \Biggr].
\label{zb}
\end{eqnarray}
For PHCC, the mass $m$ can be directly determined from the
dispersion of the $S_z=1$ excitation. We will present results
primarily in the small $|\mu|$ quantum critical region of
Fig~\ref{fig1}, although our formalism can be extended to other
regions, including across the Kosterlitz-Thouless transition into
the ``superfluid'' phase. The bare interaction between the Bose
particles, $V_0$, is renormalized by repeated interactions between
the particles, in the $T$ matrix, to the value
\begin{equation}
V_R = \frac{4 \pi \hbar^2}{m \ln (\Lambda /\sqrt{\mu^2 + (k_B
T)^2})}, \label{vr}
\end{equation}
where $\Lambda$ is a high energy cutoff, and the square-root
function in the argument of the logarithm is an arbitrary,
convenient interpolating form. The parameter $\Lambda$ is the sole
non-universal parameter appearing in the predictions of the
continuum theory. For a spin gap antiferromagnet like PHCC, we
expect $\Lambda \sim J$, where $J$ is a typical exchange constant.
For our purposes, it is convenient to rescale $V_R$ to a parameter
$U$, which has the dimensions of energy:
\begin{equation}
U \equiv \frac{2 m k_B T}{\hbar^2} V_R = \frac{8 \pi k_B T}{
(\Lambda /\sqrt{\mu^2 + (k_B T)^2})}. \label{defU}
\end{equation}

Our universal results for the continuum theory are predicated on
the assumption that the logarithm in Eqs.~(\ref{vr}) and
(\ref{defU}) is ``large''. At first glance, it would appear from
Eq.~(\ref{vr}) that the quantum theory of the Bose gas is weakly
coupled in the leading-logarithm approximation. However, as will
be clear from our analysis (and has been noted in earlier works
\cite{ss1,nikolay1,nikolay2,phase}), this is {\em not\/} the case:
although pairwise interactions are weak, the collective static and
dynamic properties of the gas remain strongly coupled even when
the logarithm  is large. We shall argue, that to leading order in
the logarithm, these strong coupling effects can be captured in an
effective classical model. The latter model is amenable to
straightforward numerical simulations, and so accurate
quantitative predictions become possible, whose precision is
limited only by the available computer time. Previous analyses
\cite{popov,fisher,dp} of the dilute Bose gas in two dimensions
assumed (when extended to the quantum critical region) that $\ln
\ln (\Lambda / (k_B T))$ was large; a more precise version of this
condition appears below, from which it is clear that this
condition is essentially impossible to satisfy in practice. We
will not make any assumption on the value of such a double
logarithm, and only make the much less restrictive assumption on a
large single logarithm.

The characteristic length and time scales of the quantum-critical
Bose gas are set by a dimensionful parameter, which we denote $R$.
Like $U$, we choose this to have dimensions of energy; so a
characteristic length is $\hbar/\sqrt{2mR}$, while a characteristic
time is $\hbar/R$. The value of $R$ is determined by the solution of
the following equation
\begin{equation}
R = -\mu + \frac{U}{2 \pi} \ln \left( \frac{\mu}{(e^{\mu/k_B T} -
1)R} \right). \label{defR}
\end{equation}
Note that this defines a $R>0$ for all $-\infty < \mu < \infty$.

\begin{figure}
\centering
\includegraphics[width=2.8in]{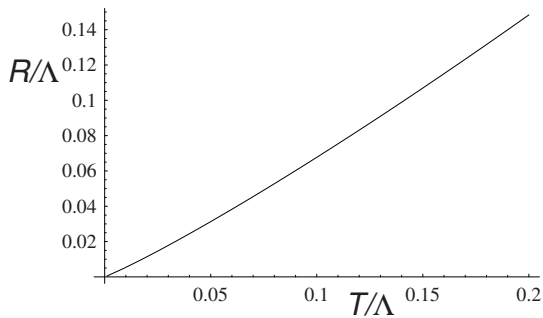}
\caption{$T$ dependence of $R$ at the quantum critical point,
$\mu=0$, obtained by solving Eq.~(\ref{defR}). $\Lambda$ is a
non-universal ultraviolet energy cutoff.} \label{rcrit}
\end{figure}

\begin{figure}
\centering
\includegraphics[width=2.8in]{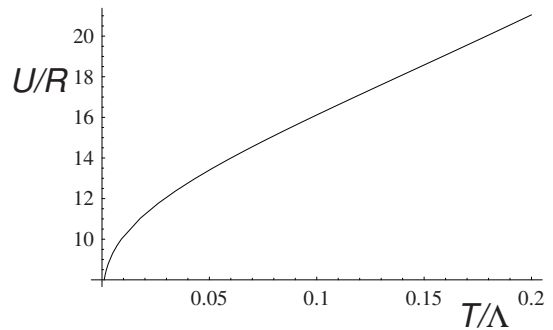}
\caption{$T$ dependence of the dimensionless ratio $U/R$ at the
quantum critical point, $\mu=0$.} \label{urcrit}
\end{figure}

To understand the magnitude of the various scales, we now discuss
approximate solutions of Eqs. (\ref{defU}) and (\ref{defR}) at the
quantum critical point, $\mu=0$. The estimates below should not be
used in place of the full solutions in applications of our results
to experiments. For the energy scale associated with $R$ we obtain
the estimate
\begin{equation}
R \sim k_B T \frac{ 4  \ln \left(\frac{1}{4} \ln (\Lambda/(k_B T)
)\right)}{\ln (\Lambda/(k_B T))}~~~;~~~\mu=0. \label{rapprox}
\end{equation}
Note that the energy scale $R$ is logarithmically smaller than
$k_B T$: this will be the key in justifying an effective classical
description of the dynamics. Numerically, we can easily go beyond
Eq.~(\ref{rapprox}), and obtain the full solution of
Eq.~(\ref{defR}) at $\mu=0$; this is shown in Fig.~\ref{rcrit}.
With the two dimensionful parameters, $U$, and $R$, at hand, the
reader will not be surprised to learn that the effective classical
theory is characterized by the dimensionless coupling $U/R$.
Indeed, as we will see, the ratio $U/R$ behaves like an effective
Ginzburg parameter for the classical theory. From
Eqs.~(\ref{defR}) and (\ref{rapprox}) we estimate
\begin{equation}
\frac{U}{R} \sim \frac{2 \pi}{ \ln \left(\frac{1}{4} \ln
(\Lambda/(k_B T) )\right)}~~~;~~~\mu=0, \label{ur}
\end{equation}
in the quantum-critical region. Again, we can go beyond the
asymptotics, and obtain precise values for $U/R$ by numerical
solution of Eqs.~(\ref{defR}) and (\ref{defU}), and the result
appears in Fig~\ref{urcrit}.
So, unless $\Lambda/(k_B T)$ is astronomically large, the ratio
$U/R$ is not small, and the classical theory is strongly coupled.
As we have already noted, we will not make the assumption of large
double logarithms here, but instead obtain numerical results for
the strongly coupled theory.

The primary result of this paper is that the {\em low energy
properties of the dilute Bose gas are universal functions of the
ratio $U/R$}. This key result is illustrated in Fig~\ref{fig1}
where we plot the loci of points with constant $U/R$, obtained
from Eqs.~(\ref{defU}) and (\ref{defR}). The physical properties
of systems along a fixed locus are the same, apart from an overall
re-scaling of energy and distance scales which are set by the
value of $R$. In most cases, the universal dependence on $U/R$ can
be determined by straightforward numerical simulations. As has
been shown earlier, in a different context\cite{ss1}, the
effective classical model undergoes a Kosterlitz Thouless
transition \cite{ss1} (see Fig~\ref{fig1}) at $U/R \approx 34$.
The values of $U/R$ in the quantum-critical region are smaller
(see Eq.~(\ref{ur}) and Fig~\ref{fig1}) and the focus of our
attention will be on these smaller values.

With an eye to neutron scattering observations in PHCC, we focus in
this paper on the frequency dependence of the boson Green's
function. In the antiferromagnet, this Green's function is
proportional to the two-point spin correlation function in the plane
orthogonal to the applied field {\em i.e.\/} the correlator of $S_+$
and $S_-$. The spectral density of this Green's function yields the
neutron scattering cross-section. In particular, we will consider
the Green's function (in imaginary time)
\begin{equation}
\chi_\psi (i \omega_n) = \frac{1}{\hbar} \int_0^{\hbar/k_B T}
\!\!\!\! d\tau \int d^2 r e^{i \omega_n \tau}  \langle \psi ({\bf
r}, \tau) \psi^\ast (0,0) \rangle; \label{defchi}
\end{equation}
after analytic continuation to real frequency, and application of
the fluctuation-dissipation theorem, we obtain the corresponding
dynamic structure factor $S_\psi (\omega)$. One of our main
results is that this structure factor obeys the scaling form
\begin{equation}
S_\psi (\omega) = \frac{k_B T}{R^2} \Phi_\psi \left( \frac{\hbar
\omega}{R}, \frac{U}{R} \right). \label{Sscale}
\end{equation}
We will numerically determine the universal function $\Phi_\psi$
here for a range of values of $U/R$ in the quantum-critical
regime.

Our results for $S_\psi (\omega)$ are obtained directly in real
time, and so do not suffer ambiguities associated with analytic
continuation. For a significant range of values of $U/R$ of
relevance to quantum criticality, we found that our numerical
results could be fit quite well with the following simple
Lorentzian functional form for $\chi_\psi (\omega)$
\begin{equation}
S_\psi (\omega) = \frac{2k_B T}{R} Z \frac{\gamma R}{(\hbar \omega
- R \omega_0)^2 + (\gamma R)^2}, \label{chires}
\end{equation}
where the scaling form in Eq.~(\ref{Sscale}) implies that the
dimensional numbers $Z$, $\omega_0$, and $\gamma$ are {\em all
universal functions of the ratio\/ $U/R$}. Notice that this
describes a neutron resonance at frequency $R \omega_0 /\hbar$
with width $R \gamma/\hbar$; our primary purpose here is to
provide theoretical predictions for the temperature dependence of
these observables. Our numerical results for the values of $Z$,
$\omega_0$ and $\gamma$ as a function of $U/R$ appear in
Figs.~\ref{zval},~\ref{omega_ur}, and~\ref{gamma_ur} later in the
paper.

\section{Quantum critical theory}
\label{sec:qc}

This section will obtain the properties of the continuum theory in
Eq.~(\ref{zb}) which were advertized above.

The analysis of the static properties of Eq.~(\ref{zb}) has been
outlined in Ref.~\onlinecite{ss1,nikolay1,nikolay2,phase}. The key
step is the integrate out all the $\omega_n \neq 0$ modes of $\psi$
to obtain an effective action only for the zero frequency component.
Among the important effects of this is to replace the bare
interaction $V_0$ by a renormalized interaction $V_R$ obtained by
summing ladder diagrams
\begin{equation}
V_R = \frac{V_0}{1 + (m V_0 /(4 \pi \hbar^2)) \ln (\Lambda/(k_B T))}
\approx \frac{4 \pi \hbar^2}{ m \ln (\Lambda/(k_B T))}.
\end{equation}
The co-efficient of $|\psi |^2$ is also renormalized, as we specify
below. The resulting effective theory for the zero frequency
component is most conveniently expressed by defining
\begin{equation}
\psi = \frac{\sqrt{2m k_B T}}{\hbar} \Psi, \label{psiscale}
\end{equation}
and rescaling spatial co-ordinates by
\begin{equation}
{\bf r} \rightarrow \frac{\hbar}{\sqrt{2m}}{\bf r}. \label{rscale}
\end{equation}
This yields the following classical partition function
\begin{eqnarray}
\mathcal{Z}_c &=& \int \mathcal{D} \Psi ({\bf r})
e^{-\mathcal{S}_c} \nonumber \\
\mathcal{S}_c &=& \int d^2 r \left[  \left| \nabla_{\bf r} \Psi
\right|^2 + \widetilde{R} |\Psi |^2 + \frac{U}{2} |\psi |^4 \right].
\end{eqnarray}
Here the energy $U$ is as defined in Eq.~(\ref{defU}), while the
`mass' $\widetilde{R}$ is given by
\begin{eqnarray}
\widetilde{R} &&= -\mu \nonumber \\ +&& \frac{2 U}{k_B T} \int
\frac{d^2 k}{4 \pi^2} \left( \frac{1}{e^{ (k^2 - \mu)/(k_B T)} - 1}
- \frac{k_B T}{ k^2 -\mu} \right) . \label{e1}
\end{eqnarray}
The most important property of this expression for $\widetilde{R}$
is that the integral of $k$ is not ultraviolet finite, and has a
logarithmic dependence on the upper cutoff. However, this is not a
cause for concern. The theory $\mathcal{Z}_c$ is itself not a
ultraviolet finite theory, and its physical properties do have a
logarithmic dependence on the upper cutoff. Fortunately (indeed,
as must be the case), the cutoff dependence in $\widetilde{R}$
above is precisely that needed to cancel the cutoff dependence in
the correlators of $\mathcal{Z}_c$ so that the final physical
results are cutoff independent. This important result is
demonstrated by noting that the only renormalization needed to
render $\mathcal{Z}_c$ finite is a `mass' renormalization from
$\widetilde{R}$ to $R$, as defined by
\begin{equation}
R  \equiv \widetilde{R} + 2 U \int \frac{d^2 k}{4 \pi^2}
\frac{1}{k^2 + R}. \label{e3}
\end{equation}
Here, the $R$ in the propagator on the r.h.s. is arbitrary, and is
chosen for convenience. We could have chosen a propagator $1/(k^2 +
c R)$ where $c$ is an arbitrary numerical constant; this would only
redefine the meaning of the intermediate parameter $R$, but not the
value of any final observable result. Combining Eq.~(\ref{e3}) with
Eq.~(\ref{e1}), we observe that the resulting expression for $R$ is
free of both ultraviolet and infrared divergences. The momentum
integrals can be evaluated, and lead finally to the expression for
$R$ already presented in Eq.~(\ref{defR}).

When expressed in terms of the renormalized parameter $R$, the
properties of the continuum theory $\mathcal{Z}_c$ are universal
({\em i.e.\/} independent of short-distance regularization). They
are defined completely by the length scale $1/\sqrt{R}$ and the
dimensionless ratio $U/R$. The field $\Psi$ is also dimensionless,
and acquires no anomalous dimension. This means {\em e.g.} the equal
time correlations obey the scaling form
\begin{equation}
S(k) = \langle | \Psi (k) |^2 \rangle = \frac{1}{R} \Phi \left(
\frac{k}{\sqrt{R}}, \frac{U}{R} \right), \label{e5}
\end{equation}
where $\Phi$ is a universal function. Note that here, and in the
remainder of the paper, we have rescaled momenta corresponding to
Eq.~(\ref{rscale})
\begin{equation}
k \rightarrow \frac{\sqrt{2m}}{\hbar} k,
\end{equation}
so that $k^2$ has the dimensions of energy. The function $\Phi$ can
be easily computed in a perturbation theory in $U/R$, and results to
order $(U/R)^2$ appear in Ref.~\onlinecite{ss1}. However, we are
interested in values of $U/R$ which are not small, and for this we
have to turn to the numerical method described in the following
subsection.

\subsection{Numerics: equal time correlations}

We will analyze numerically $\mathcal{Z}_c$ by placing it in a
square lattice of spacing $a$, and verifying that the correlations
measured in Monte Carlo simulations become $a$ independent and
universal in the limit $a \rightarrow 0$.

The partition function on the lattice is
\begin{eqnarray}
\mathcal{Z}_{cL} &=& \prod_i \int  d\Psi_i
e^{-\mathcal{S}_{cL}}  \\
\mathcal{S}_{cL} &=& \sum_{\langle i j \rangle}  \left|  \Psi_i -
\Psi_j \right|^2 + \sum_i \left[ \widetilde{R}_L a^2 |\Psi_i |^2 +
\frac{U a^2}{2} |\Psi_i |^4 \right]. \nonumber
\end{eqnarray}
The parameter $\widetilde{R}_L$ is {\em not\/} equal to the
parameter $\widetilde{R}$ above. Instead, the mapping to the quantum
theory has to be made by requiring that the values of the
renormalized $R$ are the same. In the present lattice theory we have
\begin{eqnarray}
&& \widetilde{R}_L = R \label{e10} \\
&&- 2 U \int_{-\pi}^{\pi} \frac{dk_x}{2 \pi} \int_{-\pi}^{\pi}
\frac{dk_y}{2 \pi} \frac{1}{ 4 - 2 \cos (k_x) - 2 \cos(k_y) + R
a^2}. \nonumber
\end{eqnarray}

We can measure lengths in units of $R$, and for each value of $U$,
determine $\widetilde{R}_L$ from Eq.~(\ref{e10}), and test if the
Monte Carlo correlations are independent of $a$ in the limit of
small $a$. The resulting correlations then determine the scaling
function $\Phi$ in (\ref{e5}).

\begin{figure}
\centering
\includegraphics[width=2.8in]{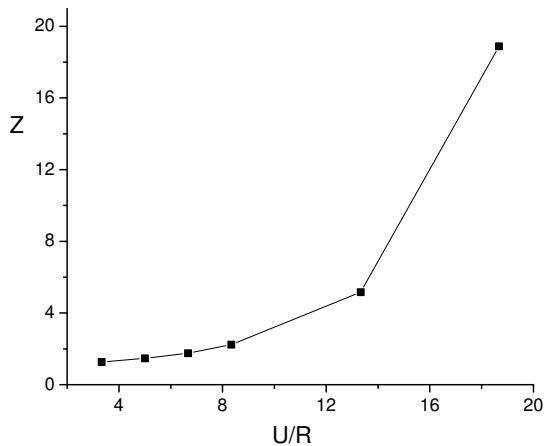}
\caption{Universal dependence of the equal-time correlation at
zero momentum on $U/R$. The parameter $Z$ is that appearing in
Eq.~(\ref{chires}).} \label{zval}
\end{figure}

We used this method to determine the values of the function $\Phi
(0, U/R)$ for a sample set of values of $U/R$ appropriate to the
quantum-critical region. Note that $\Phi(0,U/R) = Z$, where $Z$ is
the amplitude appearing in the dynamic function in
Eq.~(\ref{chires}). We used the Wolff cluster algorithm (as
described in Ref.~\onlinecite{ss1}) to sample the ensemble
specified by $\mathcal{Z}_{cL}$. Measurement of the resulting
correlations led to the results shown in Fig~\ref{zval}.

\subsection{Dynamic theory}

We now extend the classical static theory above to unequal time
correlations by a method described in some detail in
Refs.~\onlinecite{ss1} and~\onlinecite{book}. As argued there,
provided the scale $R < k_B T$, the unequal time correlations can
also be described by classical equations of motion. In the context
of perturbation theory, the reduction to classical equations of
motion is equivalent to the requirement that it is a good
approximation to replace all Bose functions by their low energy
limit:
\begin{equation}
\frac{1}{e^{\omega/T} - 1} \approx \frac{T}{\omega}.
\label{capprox}
\end{equation}
From Eq.~(\ref{rapprox}) we observe that the requirement on $R$ is
satisfied in the quantum-critical region. It also holds everywhere
in the superfluid phase, with $\mu > 0$, where $R$ becomes
exponentially small in $1/(k_B T)$ (as can be shown from
Eq.~(\ref{defR})). However, it does fail in the low $T$ `spin-gap'
region with $\mu < 0$, where $R> k_B T$. We will not address this
last region here, although a straightforward perturbative analysis
of the full quantum theory is possible here, as noted in
Ref.~\onlinecite{book}. Some results on the perturbation theory
appear in Appendix~\ref{app:pert}.

The classical equations of motion obeyed by $\Psi$ are merely the
c-number representation of the Heisenberg equations of motion obeyed
by $\psi$. With the rescalings in Eqs.~(\ref{psiscale}) and
(\ref{rscale}), these are
\begin{equation}
i \hbar  \frac{\partial \Psi}{\partial t} = \frac{\delta
\mathcal{S}_c}{\delta \Psi^\ast}.
\end{equation}
Following the reasoning leading to Eq.~(\ref{e5}), it follows that
the correlations of the $\Psi$ evolution described by these
equations of motion obey the scaling form
\begin{equation}
S(k,t) = \langle \Psi (k, t) \Psi^\ast (k,0)  \rangle = \frac{1}{R}
\Phi_t \left( \frac{k}{\sqrt{R}}, \frac{R t}{\hbar}, \frac{U}{R}
\right). \label{e7}
\end{equation}
where the scaling function $\Phi_t$ can be determined numerically,
as we describe below. After the rescalings in Eqs~(\ref{psiscale})
and (\ref{rscale}), we conclude that the Fourier transform of
$S(k, t)$ to $S(k,\omega)$ is related to the dynamic structure
factor $S_\psi (\omega)$ defined below Eq.~(\ref{defchi}) by
\begin{equation}
S_\psi (\omega) = k_B T S (0,\omega). \label{fdt}
\end{equation}

\begin{figure}
\centering
\includegraphics[width=2.6in]{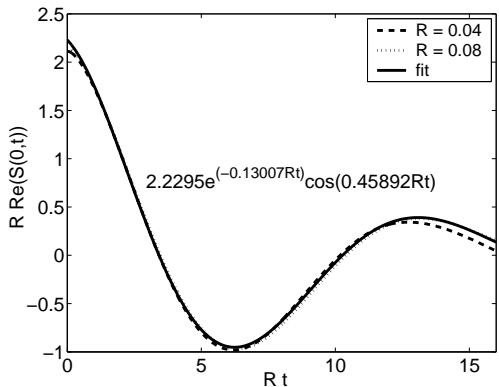}
\caption{Real part of the time-dependent structure factor $S(0,
t)$ obtained from our numerical simulation. The results above are
for $U/R=8.33$. Two different values of $R$ are shown, and the
axes have been scaled to demonstrate the scaling collapse as
required by Eq.~(\ref{e7}). The fit with the functional form in
Eq.~(\ref{sinedamp}) is also shown.} \label{time1}
\end{figure}

\begin{figure}
\centering
\includegraphics[width=2.6in]{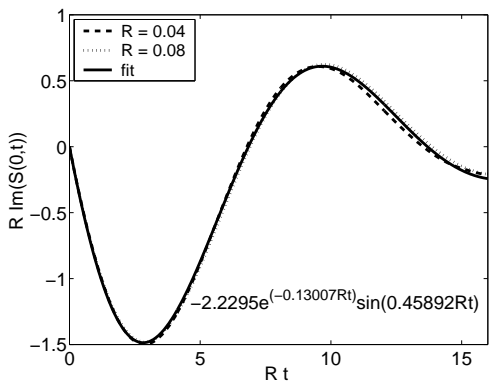}
\caption{As in Fig.~\ref{time1}, but for the imaginary part of the
correlation function.} \label{time2}
\end{figure}

We now describe our numerical computation of $\Phi_t$. We begin by
sampling the ensemble of $\Psi_i$ values specified by
$\mathcal{Z}_{cL}$, as in the previous section. Once the ensemble is
thermalized, we choose a typical set of values of $\Psi_i$ as the
initial condition. These are evolved forward in time
deterministically by solving the equations of motion
\begin{equation}
i \frac{\partial \Psi_i}{\partial \overline{t}} = \sum_{j\ n.n. i}
(\Psi_i - \Psi_j) + \widetilde{R}_L a^2 \Psi_i + U a^2 |\Psi_i |^2
\Psi_i,
\end{equation}
where $\overline{t} = t/\hbar$. For each initial condition, this
defines a $\Psi_i (\overline{t})$. Then
\begin{equation}
\Phi_t (0, R\overline{t}, U/R) = \frac{a^2 R}{N^2} \left\langle
\biggl( \sum_j \Psi_j^\ast (\overline{t} ) \biggr) \biggl(
\sum_\ell \Psi_\ell (\overline{t} ) \biggr) \right\rangle
\end{equation}
for a lattice of $N^2$ sites, and where the average is over the
ensemble of initial conditions. Sample results from such
simulations appear in Figs.~\ref{time1} and~\ref{time2}.

We have fit each observed time evolution to the functional form
\begin{equation}
\Phi_t \approx Z e^{-R (i \omega_0 \overline{t} + \gamma |
\overline{t} |)}, \label{sinedamp}
\end{equation}
where $Z$, $\omega_0$, $\gamma$ are numbers determined from the
fit. A sample fit is shown in Figs.~\ref{time1} and~\ref{time2}.
As is clear from the figures, this form provides an excellent fit
over a substantial time window. Taking the Fourier transform of
this result, and using Eq.~(\ref{fdt}), we obtain a result for
$S_\psi (\omega)$ in the form in Eq.~(\ref{chires}), with the same
parameters $Z$, $\omega_0$, $\gamma$.

\begin{figure}
\centering
\includegraphics[width=2.6in]{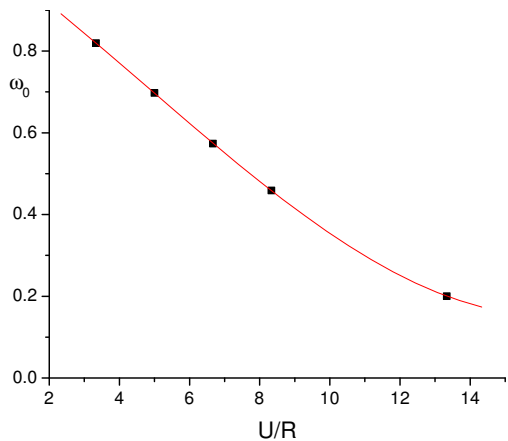}
\caption{The universal dependence of the dimensionless frequency
$\omega_0$ on $U/R$. The points are the results of the numerical
simulation, while the line is a best fit polynomial used for
interpolation.} \label{omega_ur}
\end{figure}

\begin{figure}
\centering
\includegraphics[width=2.6in]{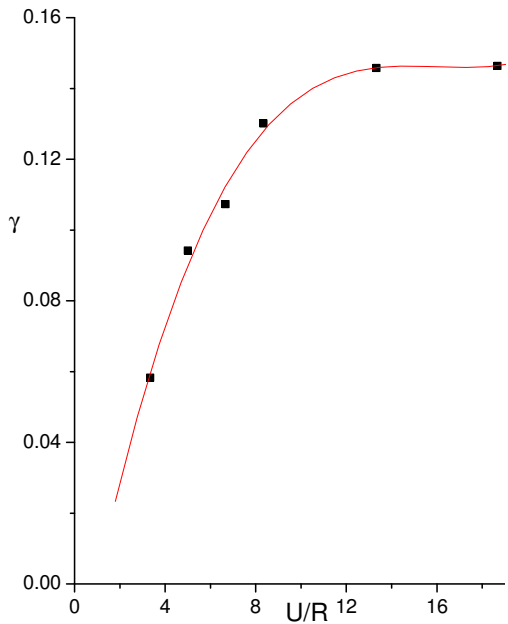}
\caption{As in Fig~\ref{omega_ur}, but for the dimensionless
damping $\gamma$. The full line at small $U/R$ is the result of
the perturbation theory obtained in Eq.~(\ref{c1}).}
\label{gamma_ur}
\end{figure}

\begin{figure}
\centering
\includegraphics[width=3.0in]{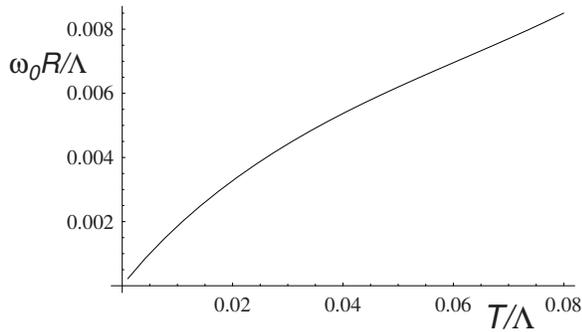}
\caption{The universal dependence of the oscillation frequency
$\omega_0 R$ on $T$ at the quantum-critical point $\mu=0$,
obtained from Figs.~\ref{urcrit} and~\ref{omega_ur}.}
\label{frequency_crit}
\end{figure}

\begin{figure}
\centering
\includegraphics[width=3.0in]{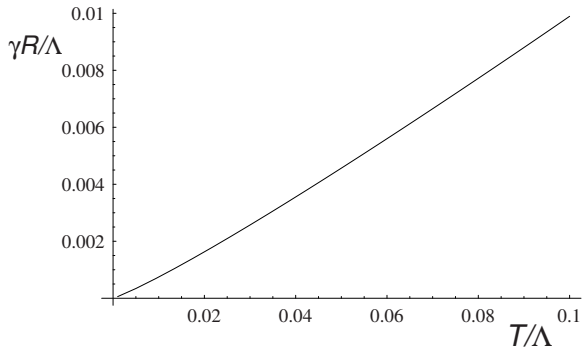}
\caption{The universal dependence of the damping rate $\gamma R$
on $T$ at the quantum-critical point $\mu=0$, obtained from
Figs.~\ref{urcrit} and~\ref{gamma_ur}.} \label{damping_crit}
\end{figure}

The results for $Z$ appeared earlier in Fig.~\ref{zval}, and our
results $\omega_0$ and $\gamma$ as a function of $U/R$ appears in
Figs.~\ref{omega_ur} and~\ref{gamma_ur}.
We can now combine the results in these figures with those in
Fig~\ref{urcrit}, and obtain predictions for the $T$ dependence of
the resonant frequeny and damping in the dynamic structure factor
in Eq.~(\ref{chires}) at the quantum-critical point. These results
appear in Fig~\ref{frequency_crit} and~\ref{damping_crit}.

\section{Conclusions}
\label{sec:conc}

This paper has argued that the quantum-critical dynamics of the
two-dimensional Bose gas (and of other quantum critical points in
two spatial dimensions) represents a strong-coupling problem.
Nevertheless, an effective classical description was obtained to
leading logarithmic order, allowing tractable numerical
simulation. The primary results of this simulation at the quantum
critical point appear in Figs.~\ref{frequency_crit}
and~\ref{damping_crit}, which specify the parameters appearing in
Eq.~(\ref{chires}).

In comparing our results to experiments, we have to set the value
of the unknown high energy cutoff $\Lambda$. For the range of
parameters shown in Figs.~\ref{frequency_crit}
and~\ref{damping_crit}, we find a roughly linear dependence of
$\omega_0 R$ and $\gamma R$ on $T$, with $\omega_0 R\approx 0.11
(k_B T)$ and $\gamma R \approx 0.10 (k_B T)$. The experiments
\cite{stone,stone2,hong} on the quantum-critical point of PHCC,
also observe a roughly linear dependence of peak frequency and
width on temperature, but with different co-efficients; the peak
frequency $\approx 0.62 k_B T$, while the width is $\approx 0.23
k_B T$. Choosing a different range of $\Lambda$ for the theory,
{\em e.g.\/} assuming the experiments are in the range $T/\Lambda
< 0.01$, will change the theoretical predictions for $\omega_0 R$
and $\gamma R$, but does not improve the agreement with
experiments.

For the peak frequency, the work of Ref.~\onlinecite{hong}
suggests an origin for the above discrepancy. A similar theory was
used in that paper to obtain the predictions of the peak
frequency, but keeping the full lattice dispersion for the spin
excitations, and the quantum Bose function values for the
occupation numbers: good agreement was found between such a theory
and the experimental observations. The continuum and classical
limits taken in the present paper were avoided.

For the width of the spin excitation, we expect that the damping
is more strongly dominated by the low energy and low momentum
excitations, and so the present continuum, classical theory should
yield a more accurate description of the experiments. This is
indeed the case, relative to the poor accuracy of the peak
frequency in the continuum theory. Nevertheless, a discrepancy of
a factor of $\approx 2$ remains between our present quantum
critical theory and the experimental observations. It is possible
that taking the classical theory also on the lattice will improve
agreement with experiments. However, there are also corrections to
the present continuum theory which could improve the situation: in
particular, at higher order in $V_R$, there appear renormalization
of the time-derivative term in Eq.~(\ref{zb}). This
renormalization would change the time-scale in the classical
equations of motion, and so change the overall frequency scale of
the results in Figs.~\ref{frequency_crit} and~\ref{damping_crit}.

\acknowledgements We thank T.~Hong and C.~Broholm for useful
discussions. This research was supported by NSF Grant DMR-0537077.

\appendix

\section{Pertubation theory}
\label{app:pert}

Here we present the results of a direct perturbative computation of
$\chi_\psi$. To order $U^2$, perturbation theory yields\cite{ss1}
\begin{equation}
\chi_\psi (\omega) = \frac{1}{- \hbar \omega + R + \Sigma (\omega)}
\label{chipert}
\end{equation}
where
\begin{widetext}
\begin{eqnarray}
\Sigma (i \omega) = - 2 U^2 \sum_{\epsilon_1, \epsilon_2} \int
\frac{d^2 k_1}{4 \pi^2} \frac{d^2 k_2}{4 \pi^2} \frac{1}{(-i
\epsilon_1 + k_1^2  + R)(-i \epsilon_2 + k_2^2 + R)( -i
(\epsilon_1 + \epsilon_2) + i \omega + (k_1 + k_2)^2 + R )}.
\nonumber
\end{eqnarray}
The frequency summation is done most easily by partial fractions and
yields
\begin{eqnarray}
\Sigma ( i \omega) &=& - 2 \left( \frac{U}{k_B T} \right)^2  \int
\frac{d^2 k_1}{4 \pi^2} \frac{d^2 k_2}{4 \pi^2} \frac{\left[
n(k_1^2 +R ) - n((k_1 + k_2)^2 +R) \right] \left[n(k_2^2 +R) -
n((k_1 + k_2)^2 - k_1^2)\right]}{ i \omega + (k_1 + k_2)^2 - k_1^2
- k_2^2 - R}. \nonumber
\end{eqnarray}
Now we analytically continue to real frequencies, and take the
imaginary part at $\omega = R/\hbar$, which is the leading order
position of the pole in Eq.~(\ref{chipert}). This yields
\begin{eqnarray}
\mbox{Im} \Sigma (R/\hbar) &=& 2 \pi \left( \frac{U}{k_B T}
\right)^2 \int \frac{d^2 k_1}{4 \pi^2} \frac{d^2 k_2}{4 \pi^2}
\left[ n(k_1^2 +R ) - n((k_1 + k_2)^2 +R) \right]
\nonumber \\
&~&~~~~~~~~\times\left[n(k_2^2 +R) - n((k_1 + k_2)^2 -
k_1^2)\right]  \delta( (k_1 + k_2)^2 - k_1^2 - k_2^2) \nonumber \\
&=& - 2 \pi \left( \frac{U}{k_B T} \right)^2 \int \frac{d^2 k_1}{4
\pi^2} \frac{d^2 k_2}{4 \pi^2} \left[ n(k_1^2 +R ) - n(k_1^2 + k_2^2
+R) \right]
\left[n(k_2^2) - n(k_2^2 + R)\right]  \nonumber \\
&~&~~~~~~~~~~~~~~~~~~~~~~~~~~~~~~~~~~\times \delta( (k_1 + k_2)^2
- k_1^2 - k_2^2).
\end{eqnarray}
Now we can do the angular integration because the angle appears only
in the argument of the delta function, and obtain
\begin{eqnarray}
\mbox{Im} \Sigma (R/\hbar) &=&  -  \left(\frac{U}{ 2 \pi k_B
T}\right)^2 \int_0^\infty \! dk_1 \int_0^\infty \! dk_2 \frac{
(e^{R/(k_B T)} - 1)}{(e^{(k_1^2 + R)/(k_B T)}-1) (e^{(k_2^2 +
R)/(k_B T)}-1) ( 1 - e^{-(k_1^2 + k_2^2 + R)/(k_B T)}) }.
\label{q1}
\end{eqnarray}
This result is a function of $R/(k_B T)$ which has to be evaluated
numerically. However, it is instructive to examine its value in the
classical limit, upon applying the approximation in
Eq.~(\ref{capprox}), when we obtain
\begin{eqnarray}
\mbox{Im} \Sigma (R/\hbar) &=&  -  \frac{U^2 R}{4 \pi^2 }
\int_0^\infty \! dk_1 \int_0^\infty \! dk_2 \frac{1}{(k_1^2 +
R)(k_2^2 + R)(k_1^2 + k_2^2 + R)} \nonumber \\ &=& - R \left(
\frac{U}{R} \right)^2 \frac{(4-\pi)}{16 \pi}. \label{c1}
\end{eqnarray}
Notice that explicit factors of $k_B T$ have dropped out, and
consequently Eqs.~(\ref{chipert}) and (\ref{c1}) are consistent
with the scaling form Eq.~(\ref{chires}). It is also interesting
to compare the value of the classical limit in Eq.~(\ref{c1}) with
that obtained from the full quantum expression in Eq.~(\ref{q1}).
For $R=k_B T/2$ (which are roughly the quasiparticle energy values
obtained in the experiments on PHCC \cite{hong}), Eq.~(\ref{q1})
evaluates to a value which is 16\% smaller than Eq.~(\ref{c1}).
This gives us an estimate of the error made by representing the
quantum-critical theory by classical equations of motion.
\end{widetext}


\end{document}